\documentclass[acmlarge,nonacm]{acmart}

\AtBeginDocument{%
  }

\acmJournal{DTRAP}

\begin{document}

\title{Investigating Threats Posed by SMS Origin Spoofing to IoT Devices}

\author{Akaki Tsunoda}
\email{research@akaki.io}
\orcid{0000-0002-1114-7986}
\affiliation{
  \institution{Independent Researcher}
  \country{Japan}
}


\begin{abstract}
The short message service (SMS) is a service for exchanging texts via mobile networks that has been developed not only as a means of text communication between subscribers but also as a means to remotely manage Internet of Things (IoT) devices. However, the originating number of an SMS can be spoofed. If IoT devices authenticate administrators based on the originating number of an SMS, the authentication is bypassed via SMS origin spoofing. Consequently, IoT devices are at risk of accepting commands from attackers and performing unauthorized actions. Accordingly, in this study, the specifications of major cellular IoT gateways were evaluated by focusing on remote management via SMS, and the authentication bypass hypothesis was verified. The results showed that 25 of the 32 targeted products supported SMS-based remote management, and 20 implemented authentication based on the originating number of the SMS. Furthermore, by spoofing the originating number of the SMS, one product was demonstrated to be remotely exploitable through authentication bypassing. Thus, this study revealed the threats posed by SMS origin spoofing to IoT devices and proved that SMS origin spoofing not only threatens text communication between people but also puts machine communication at risk.
\end{abstract}





\maketitle

\section{Introduction}
The short message service (SMS) is used for remotely managing Internet of Things (IoT) devices. Originally, the SMS was developed for mobile subscribers to exchange text messages \cite{sms}. The capability of SMS to send and receive text over long-range wireless communications can also be utilized for communication with machines, such as IoT devices \cite{pistek}. IoT devices on which this study focuses are administered by servers over the network for various controls, such as rebooting them or changing their settings. These devices are actively used as sensors to detect crop growth and weather conditions in farmlands, or vibrations in buildings, bridges, and other structures, and as tracking devices for delivery trucks \cite{xu, mishra2022, ding}. Data observed from the real world using IoT devices are collected and analyzed by an administration server to create new values. In these cases, most IoT devices are installed in remote locations that are physically inaccessible to administrators. Thus, in many cases, short-range wireless communication cannot be used to access IoT devices, and it is not economically feasible to develop a new wired network to solve the problem. IoT devices installed in remote locations can be accessed using existing SMS technology, as long as they are within the coverage area of the mobile network. Thus, the remote management of IoT devices can be facilitated by delivering commands to the devices via SMS.

However, the sender information in an SMS can be spoofed by attackers. The sender information of an SMS is specified by the sender’s phone number (henceforth referred to as the “originating number”) or a display name called the “sender ID” composed of alphanumeric characters. The sender ID can be specified as any string using the SMS gateway service. Thus, it can be spoofed as the name of a trusted entity \cite{akaki}. However, short messages with specified sender IDs cannot receive a reply and are unavailable in some countries \cite{smsto}. Therefore, the originating number is generally used as the sender information to enable two-way communication between subscribers or machines via SMS. Previous studies have shown that originating numbers can also be spoofed as arbitrary phone numbers \cite{tsunoda, tu, yu, kim}. Short messages with spoofed sender information have been exploited in various cybercrimes, such as phishing scams \cite{royalmail, posb, citizenlab}. For example, an attacker can impersonate a widely trusted entity and send a message including the URL of a website that requires the victim to enter authentication or payment information. Unfortunately, because the sender’s information displayed in short messages cannot be trusted, detection and mitigation measures for such attacks have become crucial research topics \cite{mishra2020, jain, joo, akande}.

If IoT devices authenticate the sender of an SMS command based on the originating number, the authentication can be bypassed via SMS origin spoofing. An IoT device that receives a remote-management command should authenticate that the sender of the command is an authorized administrator. Otherwise, the IoT device is likely to execute malicious commands sent by an attacker, potentially harming the concerned parties. For example, if the attacked IoT device is a sensor, factory production could be disrupted by an alarm or a machine shutdown triggered by false readings. In the case of a tracking device, false or sensitive location information, such as details about personal or valuable items, could be divulged. Therefore, authentication is an essential for preventing such unauthorized remote operations on IoT devices. Based on the SMS specification, only subscribers to a specific phone number can send a short message from that number, and thus, the identity of the subscriber can be authenticated based on the ability to send. This authentication method uses the originating number of the short message as the credential and verifies that the number matches the subscriber's phone number. However, when an IoT device authenticates the sender of a command received via SMS using this method, authentication can be bypassed via SMS origin spoofing.

Based on this hypothesis, this study investigates the threat posed by SMS spoofing to IoT devices. Previous studies on SMS origin spoofing have only considered threats to mobile phone users and not to IoT devices \cite{tsunoda, tu, yu, kim}. In recent years, studies concerning the security of existing IoT implementations have increased \cite{rajmohan}, and the security of IoT devices has also become a crucial research topic \cite{song, aziz, bella, færøy}. Furthermore, several studies have focused on threats posed by remote attacks on IoT devices \cite{alrawi, butun, kuang}. Hussain et al. [2016] and Pistek et al. [2020] mentioned the threats posed by SMS spoofing on IoT devices \cite{hussain, pistek}. However, they did not conduct a detailed investigation focused on these threats, nor did they attempt to verify the hypothesis. To fill this gap in previous studies, authentication methods for SMS-based remote management implemented in IoT devices are investigated in this study and the bypassability of authentication based on the originating number of an SMS (henceforth referred to as “originating number-based authentication”) is demonstrated. Consequently, the threat posed by SMS origin spoofing to IoT devices is revealed and threat mitigation measures are devised.

\section{Background}

This chapter describes the remote management of IoT devices via SMS and the threats posed when authentication protection is bypassed. Unlike traditional Internet devices that are installed indoors or carried by users at all times, IoT devices, such as sensors or tracking devices, are installed outdoors, typically in hard-to-reach locations, or in multiple unspecified locations. In such cases, the administrator cannot visit all installations or connect them using wired cables to manage the devices. This problem is solved by using existing mobile network technologies. The following sections present the advantages of using SMS and related specifications, and then describe the threats posed to IoT devices by authentication bypassing based on SMS origin spoofing.

\subsection{Remote Management Using SMS}

The SMS is used for the remote management of IoT devices because it is economically feasible in several ways. SMS technology has been deployed in mobile networks, from 2G to 5G; thus, it is compatible with radio access networks provided by mobile network operators (MNOs) worldwide. In other words, SMS achieves near-universal coverage. Moreover, as SMS technology does not require a permanent Internet connection, it can reduce power consumption \cite{squire}. This is a significant advantage for power-limited IoT devices, such as those powered by batteries, that must operate for long periods. Additionally, since SMS has been deployed in the latest 5G mobile network, it will continue to be available in the foreseeable future. The Global System for Mobile Communications Association (GSMA), an international nonprofit organization in the mobile communication industry, expects the number of IoT devices connected to 2G to 5G mobile networks to grow from 694 million in 2018 to 1.3 billion by 2025 \cite{gsma}. Therefore, SMS is expected to be used for the remote management of IoT devices.

Several international organizations have proposed SMS-based remote management mechanisms. The 3rd Generation Partnership Project (3GPP), an international standards organization for mobile communication technologies, defines machine-to-machine (M2M) communication between IoT devices or their administration servers as machine-type communications (MTC); its architecture is considered in TR 23.888 \cite{mtc}. TR 23.888 considers an architecture that uses SMS as a triggering mechanism for MTC. OneM2M, a global partnership project for IoT/M2M device interoperability, uses SMS as a connection method between IoT devices and mobile networks. Specifically, oneM2M uses Open Mobile Alliance (OMA) Device Management (DM) as its SMS-based device management method \cite{onem2m}. The OMA DM is a protocol for device management developed by OMA, a standards organization for mobile phone technology, which provides a mechanism for sending device management notifications via SMS \cite{omadm}. This implies that SMS is recognized by international organizations as effective for the remote management of IoT devices.

\subsection{Threat Model}

When originating number-based authentication is used in SMS-based remote management, a fake message that spoofs the origin to a legitimate phone number can bypass the authentication. IoT devices must implement sender authentication to avoid accepting illegitimate SMS commands from third parties without administrative privileges. A possible authentication method involves verifying that the originating number of an SMS matches the phone number of the pre-registered administrator to identify legitimate SMS commands sent from the administrator’s mobile phone or server. However, the originating number of an SMS can be spoofed as another person’s phone number by exploiting the capability of the relevant protocol \cite{tsunoda, tu} or vulnerabilities in the protocol \cite{yu, kim}. Accordingly, attackers can bypass authentication using fake messages that specify the administrator’s phone number registered on the IoT device as the origin.

Once IoT devices have been bypassed for authentication, they can be remotely controlled by an attacker and exploited in various types of attacks. Threats from authentication bypasses depend on the actions that the IoT device allows via remote management. Two abstract actions are given below as examples, including the threats if they are exploited, divided by the attack channel.

\textbf{State change action.} If the actions of stopping an IoT device or changing its settings are exploited, threats such as denial-of-service (DoS) attacks and the destruction or falsification of recorded data are expected to impact the IoT device itself. Additionally, threats can affect other devices connected to the IoT device and its direct users.

\textbf{State retrieve action.} If the actions of sending the IoT device status or collected data to the administrator are exploited, threats can impact the administrator of the IoT device. For instance, sending incorrect data through fake messages can disrupt the functionality of the administration server or generate reports based on false information. Additionally, an attacker can steal sensitive information stored on IoT devices, such as authentication credentials or useful data collected by sensors.

\section{Methods}

This chapter describes the methods for a fact-finding survey of authentication based on the SMS origin in IoT devices and the verification of its bypassability. This study focuses on cellular IoT gateways as the investigated IoT devices. A cellular IoT gateway is a general-purpose IoT device that controls network traffic when an application-specific IoT device, such as a sensor, accesses an administration server over the Internet from within a local network. It enables control via wireless connections over mobile networks in environments where wired cables cannot reach or when mission-critical systems require redundancy. Based on data from 2021, the value of the global cellular IoT gateway market is increasing by 14\% year-on-year, and this growth trend is expected to continue \cite{berg}. Therefore, this study conducted a fact-finding survey and hypothesis verification on cellular IoT gateways, which are expected to have high versatility and potential. The following sections first describe how the target products are selected and how the fact-finding survey is conducted on them, and then present the verification process for the bypassability of authentication.

\subsection{Fact-Finding Survey of Originating Number-Based Authentication}

A fact-finding survey was conducted on products from major cellular IoT gateway vendors. The survey covered existing products from 32 vendors listed in a market report on cellular IoT gateways published by Berg Insight in 2022 \cite{berg}. The detailed features of SMS-based remote management were examined based on the documentation available online for these products. Specifically, the availabilities of SMS-based remote management and originating number-based authentication, as well as the requirements for authentication for each product, were reviewed, based on which the impact of SMS origin spoofing was assessed. Although the category name of the target product slightly varies from vendor to vendor, such as “industrial IoT gateway” or “cellular router,” this study uses the name “cellular IoT gateway” in accordance with this market report.

\subsection{Verification of Authentication Bypass via SMS Origin Spoofing}

\begin{figure}
  \centering
  \includegraphics[width=0.65\linewidth]{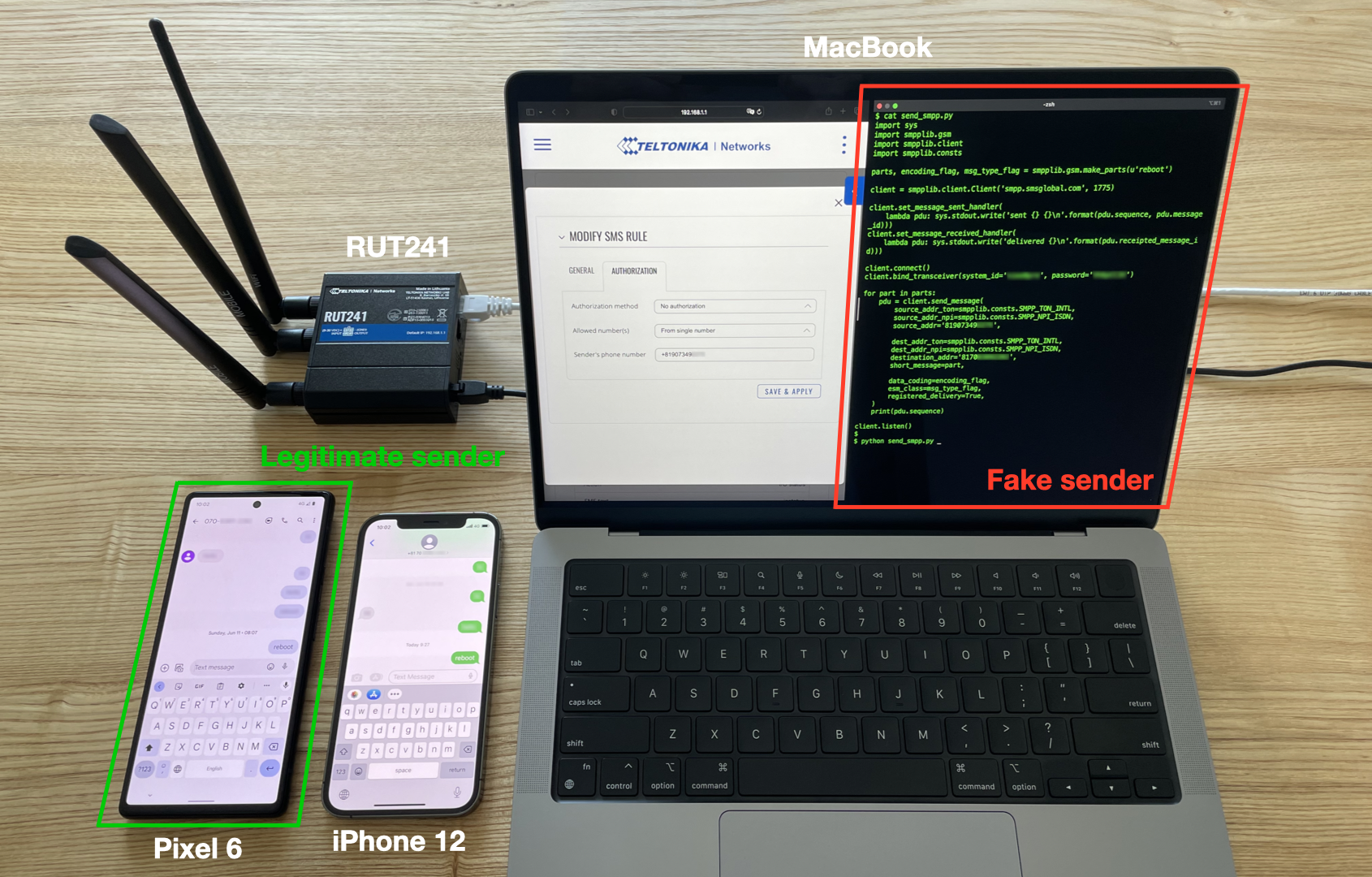}
  \caption{All devices used in the validation process.}
  \label{fig:devices}
\end{figure}

A cellular IoT gateway that implements originating number-based authentication was used to verify the hypothesis of the threat posed by SMS origin spoofing to IoT devices. Figure \ref{fig:devices} shows all the devices used in the validation of this study. A RUT241 device from Teltonika Networks was selected as the gateway product for verification because it has been approved for use in Japan, where the verification was conducted \cite{rut241}. A SIM card provided by a Japanese MNO was inserted into the RUT241 device, and a provided phone number was assigned. In addition, their SIM cards were inserted into a Pixel 6 smartphone that simulated the administrator’s mobile phone, and iPhone 12 that simulated the non-administrator’s mobile phone. Moreover, the phone number assigned to the Pixel 6 was registered in the authentication settings for SMS-based remote management of the RUT241. Verification was conducted using the latest firmware, RUT2M\_R\_00.07.04.3, available for the RUT series at the time of verification. In this firmware, all SMS-based remote management actions and password-based authentication are enabled by default. For verification purposes, password-based authentication was disabled, leaving only originating number-based authentication active.

To spoof the originating number of an SMS, Short Message Peer-to-Peer (SMPP), a protocol for forwarding short messages, was used. Tsunoda [2024] showed that the SMPP protocol can be used to spoof the originating number of an SMS as another person’s phone number on a Japanese mobile network \cite{tsunoda}. This study reused the Python script used in his study to send an SMPP message forwarding a fake SMS command with a spoofed originating number. The phone number of the Pixel 6, that is, the phone number registered in the authentication settings of RUT241, was specified in the source\_addr field of the SMPP message, and the phone number of the RUT241 was specified in the destination\_addr field. The SMS command “reboot” was specified in the text of the SMPP message to trigger the reboot action.

\begin{figure}
  \centering
  \includegraphics[width=0.85\linewidth]{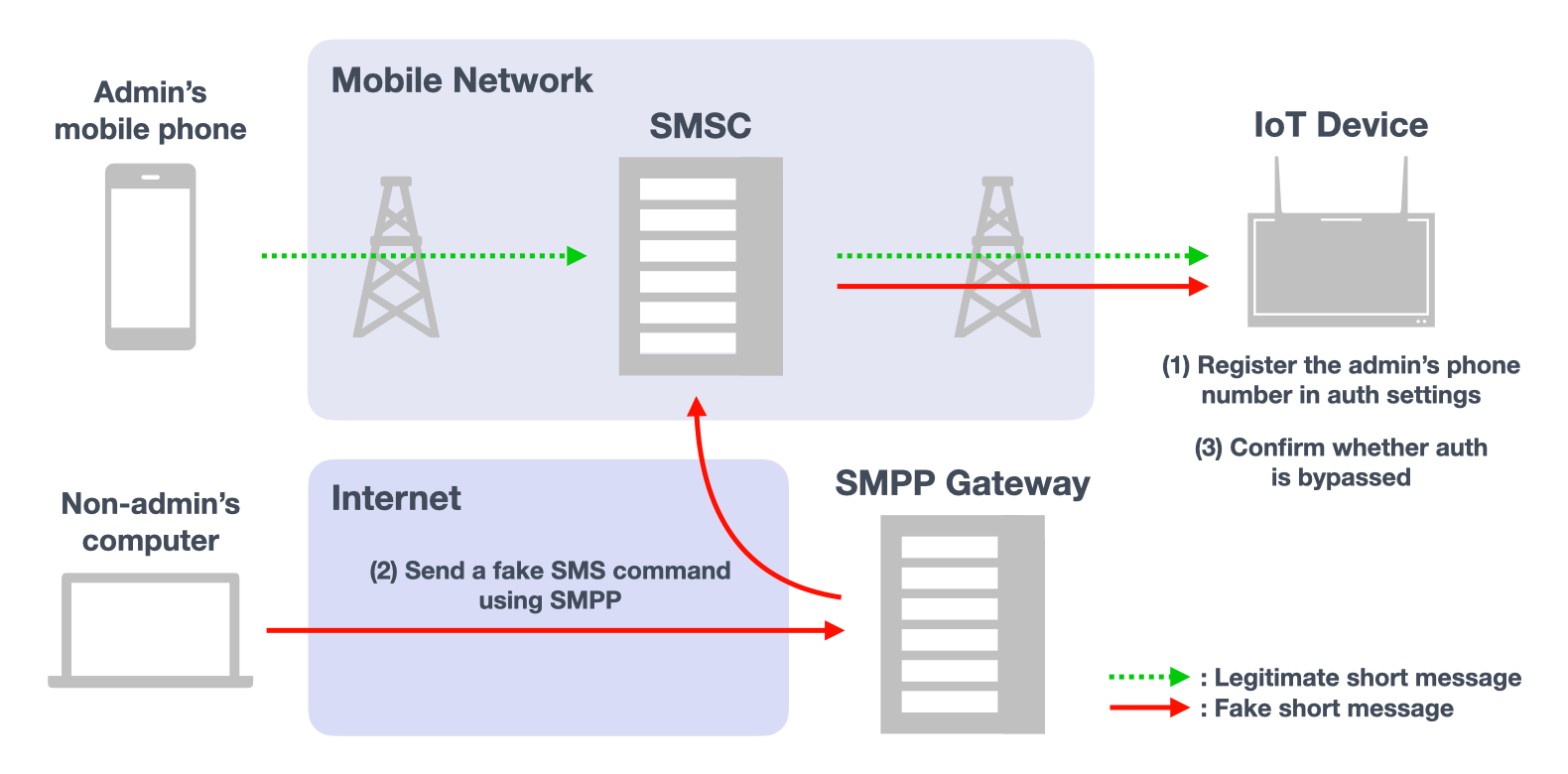}
  \caption{Overall verification process of authentication bypass via SMS origin spoofing. (1) Registering the administrator’s phone number in the authentication settings; (2) sending a fake SMS command using SMPP; (3) confirming whether originating number-based authentication is bypassed.}
  \label{fig:overall}
\end{figure}

It was confirmed whether a fake SMS command with a spoofed originating number using SMPP could bypass originating number-based authentication. The overall verification process is illustrated in Figure \ref{fig:overall}. The verification started by registering the administrator’s phone number in the originating number-based authentication settings of the RUT241 device. Next, the Python script, which was executed on a MacBook that simulated the non-administrator’s computer, submitted a command to the SMPP gateway using a fake short message with the origin spoofed as the administrator’s phone number. The command was then delivered to the RUT241 through the Japanese mobile network. Finally, the RUT241 was monitored to verify if the device received the fake short message with a spoofed originating number, if the command bypassed the originating number-based authentication of the RUT241, and if, consequently, the device rebooted as commanded. The RUT241 logs generated by the verification were analyzed, confirming whether the originator number of the received fake SMS command matched the legitimate one.

\section{Results}

This chapter reports the results of the fact-finding survey of authentication based on SMS origin in IoT devices and the verification of its bypassability. In this study, the originating number-based authentication for the products of major cellular IoT gateway vendors was surveyed. The results show that 25 of the 32 targeted products support SMS-based remote management, and 20 of them implement originating number-based authentication. Moreover, by using a fake message with a spoofed originating number, it is demonstrated that in one of the products, authentication could be bypassed to execute remote management commands. The following sections first present the detailed results of the fact-finding survey, and then describe the results of the hypothesis verification.

\subsection{Originating Number-Based Authentication}

Of the 32 surveyed products, 25 support SMS-based remote management. Table \ref{tab:details} presents the survey results for each product. The datasheets or user manuals for the 25 products describe the capability to change or retrieve some states of the device, triggered by a short message in a defined text format. Some products allow various actions with multiple SMS commands, whereas others only allow device reboots; however, in Table \ref{tab:details}, both conditions are indistinctively presented as SMS-based remote management, as the primary focus of this survey was to clarify the current status of originating number-based authentication. Products that do not support SMS-based remote management rely on remote management through IP-based protocols over mobile networks.

Of the 25 products that support SMS-based remote management, 20 implemented originating number-based authentication. The user manuals or OS configuration guides for the 20 products describe the existence of a configuration item in the SMS-based remote management settings that only accepts SMS commands received from registered phone numbers. Of the 20 products, several implement password-based authentication in addition to originating number-based authentication, and four require it. Three products have a mechanism that allows originating number-based authentication, depending on the implementation of user scripts using the vendor’s own software development kit (SDK) or OS commands. Finally, the remaining two products support SMS-based remote management according to their datasheets, but owing to an information deficit caused by the unavailability of a user manual, it could not be determined whether these products support originating number-based authentication. Requests for user manuals were made to these two vendors; however, the expected responses were not forthcoming.

\begin{table}
  \caption{SMS-based remote management details in major cellular IoT gateways.}
  \label{tab:details}
  \begin{tabular}{p{0.25\textwidth}p{0.20\textwidth}p{0.165\textwidth}p{0.285\textwidth}}
    \toprule
    \multicolumn{1}{c}{Vendor name} &
    \multicolumn{1}{c}{Product name} &
    \multicolumn{1}{c}{\begin{tabular}[c]{@{}c@{}}SMS-based remote\\management\end{tabular}} &
    \multicolumn{1}{c}{\begin{tabular}[c]{@{}c@{}}Originating number-based\\authentication\end{tabular}} \\
    \midrule
    Acksys & AirBox LTE & No & \:\:- \\
    ADLINK Technology & MXE-210 & No & \:\:- \\
    Advantech & ICR-2031 & Yes & Yes \\
    Belden & OWL LTE & Yes & Yes \\
    Casa Systems & NTC-220 & Yes & Yes (Password required) \\
    Cisco & ISR 1100 & Yes & Yes (Implementation dependent) \\
    Cradlepoint (Ericsson) & E3000 & Yes & Yes (Password required) \\
    Digi International & IX30 & Yes & Yes (Implementation dependent) \\
    Eurotech & ReliaGATE 10-14 & No & \:\:- \\
    Four-Faith & F2X64 & Yes & No \\
    HMS Networks & Ewon Cosy+ & Yes & No \\
    Hongdian & H8922 & Yes & Yes \\
    InHand Networks & InRouter305 & Yes & Yes \\
    INSYS Microelectronics & MIRO & Yes & Yes \\
    Lantronix & X300 & Yes & Yes \\
    Matrix Electrónica & MTX-Router-Titan II-S & Yes & Yes \\
    MB Connect Line (Red Lion) & mbNET RA70A & Yes & Yes \\
    MC Technologies & MC MRLQ2 & Yes & No \\
    Milesight & UR32L & Yes & Yes \\
    Moxa & OnCell 3120 & Yes & Yes \\
    MultiTech & Cell 100 & Yes & Yes \\
    NetModule (Belden) & NB1601-La & Yes & Yes (Implementation dependent) \\
    Option (Crescent) & CloudGate LTE WW & No & \:\:- \\
    Peplink & BR1 Mini & Yes & Yes (Password required) \\
    Queclink Wireless Solutions & WR201LG & Yes & Information deficit \\
    RAD & SecFlow-1v & Yes & Information deficit \\
    Red Lion & RAM-6021 & No & \:\:- \\
    Robustel & R2011 & Yes & Yes (Password required) \\
    Sagemcom Dr. Neuhaus & TAINY IQ-LTE & No & \:\:- \\
    Systech Corporation & SysLINK SL-500 & No & \:\:- \\
    Teltonika Networks & RUT241 & Yes & Yes \\
    Westermo & MRD-405 & Yes & Yes \\
    \bottomrule
  \end{tabular}
\end{table}

\subsection{Authentication Bypass Demonstration}

Before proceeding to the hypothesis verification, the normal operation of the originating number-based authentication in the cellular IoT gateway was confirmed. This stage involved validating if, once the phone number assigned to the Pixel 6 was registered in the originating number-based authentication settings of the RUT241, only SMS commands received from that phone number were accepted. A short message with the text “reboot” was sent from the Pixel 6 smartphone to the phone number assigned to the RUT241. As the phone number of the Pixel 6 is allowed in the authentication settings, the message was accepted as an SMS command and the RUT241 was rebooted. In contrast, a message with the same text sent from the iPhone 12 was rejected by the RUT241 because the phone number of the iPhone 12 was not allowed in the authentication settings; hence, it did not reboot. These differences in behavior indicate normal operation of originating number-based authentication implemented in the RUT241 device.

\begin{figure}
  \centering
  \includegraphics[width=0.85\linewidth]{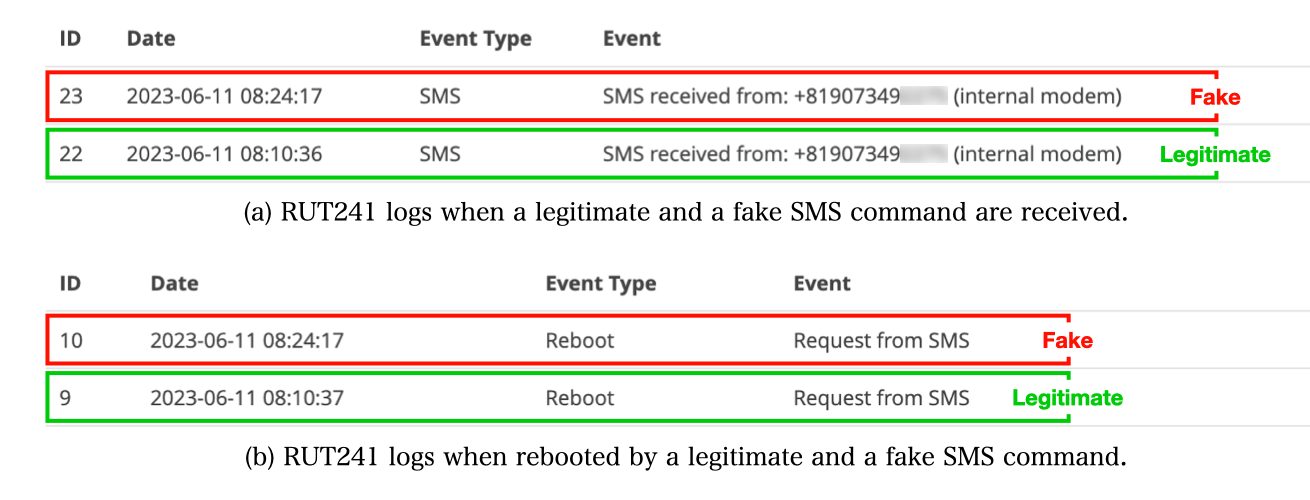}
  \caption{RUT241 logs in response to a legitimate and a fake SMS command.}
  \label{fig:logs}
\end{figure}

The originating number-based authentication implemented in the cellular IoT gateway was bypassed using an SMPP message with a spoofed origin. Specifically, the SMPP message loading the text “reboot” was sent using the Python script executed by the non-administrator, and the RUT241 that received the text was confirmed to reboot. Figure \ref{fig:logs} shows the RUT241 log output upon receiving a legitimate SMS command and a fake command with a spoofed origin. As shown in Figure \ref{fig:logs} (a), the originating numbers of the two SMS commands match. Furthermore, as shown in Figure \ref{fig:logs} (b), the fake SMS command with a spoofed origin was accepted by the RUT241, and the reboot action was executed. This proves the hypothesis of this study that originating number-based authentication for SMS-based remote management can be bypassed via SMS origin spoofing.

\section{Discussion}

This chapter discusses the threat posed by SMS origin spoofing to IoT devices based on the results of the fact-finding survey and hypothesis verification. In this study, the SMS origin spoofing proposed in previous studies is extended to demonstrate that it can be used to bypass authentication implemented in IoT devices. As mentioned before, previous research has shown that the originating number of an SMS can be spoofed by exploiting the capability and vulnerabilities of the relevant protocols. However, while previous studies have focused on the threats posed to smartphone users, this study focused on the threats posed to IoT devices. The results of this study prove that the SMS origin spoofing demonstrated in previous research puts machine communication at risk. The following sections first devise a threat scenario posed by SMS origin spoofing to cellular IoT gateways, discuss threat mitigation measures, and finally provide recommendations for future studies.

\subsection{Threat Scenario}

\begin{figure}
  \centering
  \includegraphics[width=0.65\linewidth]{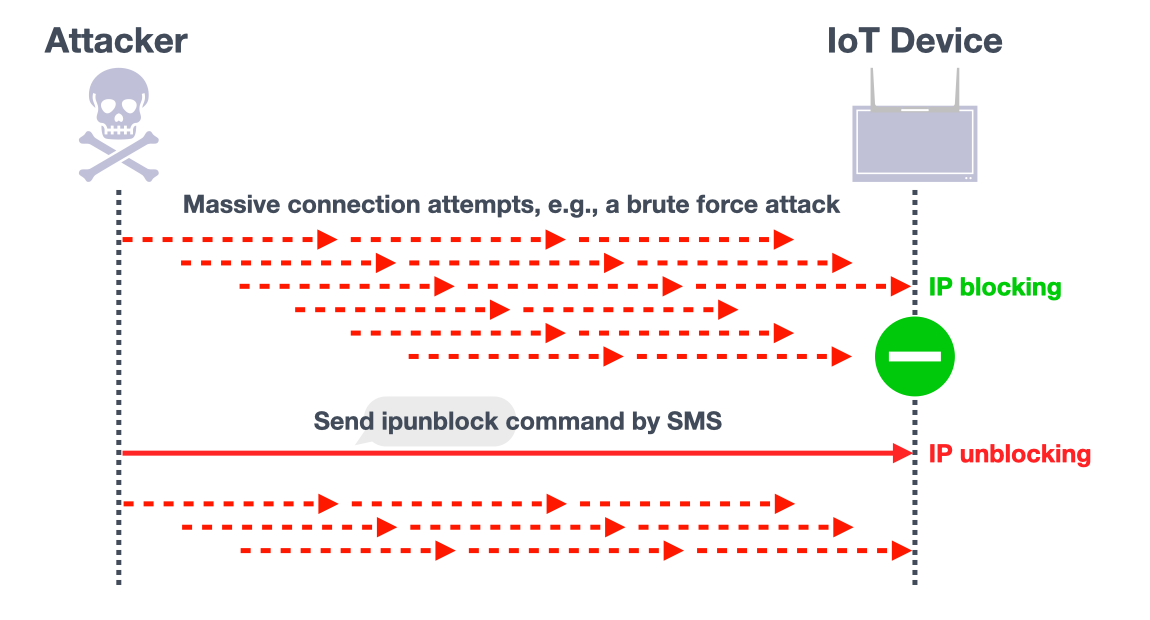}
  \caption{Threat scenario that exploits the SMS command to disable IP blocking.}
  \label{fig:threat}
\end{figure}

The severity of a threat after the authentication of an IoT device is bypassed depends on the actions allowed by the SMS-based remote management. The threats from exploiting the action of rebooting an IoT device are limited to a temporary service outage. In contrast, exploiting an action to change the settings of protection features of a device can be serious. For example, the SMS remote management of the RUT241 verified in this study allows an action to disable the feature that blocks remote access based on IP address. IP blocking is a protection feature that is activated when the limit of failed connections to the Web or Secure Shell (SSH) is exceeded and denies access requests from the IP address of the connection source. However, the SMS remote management of the RUT241 triggers the SMS command “ipunblock” to unblock the blocked IP address \cite{rut241wiki}. Therefore, attackers can continue their attack unaffected by IP blocking by using this SMS command, as shown in Figure \ref{fig:threat}. Specifically, an attacker attempting to remotely log in to an IoT device via SSH will be denied access owing to a password-based authentication failure that exceeds the limit. Nevertheless, by sending this SMS command, the attacker can unblock himself/herself and resume the authentication attempts. If the SSH login is breached, the attacker could take complete control of the device.

Moreover, some actions allowed by SMS-based remote management could escalate into various threats. Several cellular IoT gateways allow the use of SMS to execute AT commands \cite{x300, mtx, mbnet} or download firmware updates from arbitrary servers \cite{oncell}. If such a device is attacked, a greater impact on the network to which it connects can be expected. For example, because AT commands have capabilities that extend beyond modem control, attackers can execute arbitrary commands to bypass security mechanisms or steal sensitive information \cite{tian}. Similarly, if the firmware is updated to a malicious version created by an attacker, the device can be hijacked and used as a springboard for network intrusion or remote tampering \cite{bettayeb}. As the actions allowed by SMS-based remote management become more advanced, threats become increasingly serious when authentication is bypassed.

However, for such threat scenarios to occur in the real world, two conditions must be satisfied. First, the attacker must know in advance both the phone number assigned to the target device and that to be spoofed as the origin, i.e., the phone number registered in the authentication settings of the target device. In this study, the phone number to be spoofed was first pre-determined, and verification was subsequently initiated. In the real world, attackers must obtain it through other channels, such as open-source intelligence or social engineering. Second, the mobile networks to which the target device is connected should be vulnerable to SMS origin spoofing. As Tsunoda [2024] showed, the spoofability of the originating phone number of an SMS depends on the MNO’s delivery technology or security policies. In this study, the hypothesis was verified using the vulnerability discovered in the previous study. However, in an actual attack, attackers must discover and exploit vulnerabilities in the MNOs used by the victims. In summary, the feasibility of the threats pointed out in this study depends on the knowledge of the attacker and the mobile network to which the IoT device is connected.

\subsection{Mitigation Measures}

The results of this study suggest that supporting authentication that is not based on the origin of the SMS could serve as a measure to mitigate authentication bypass via SMS origin spoofing. Several IoT devices support password-based authentication for SMS-based remote management, but this still leaves the risk associated with using clear text passwords, i.e., the possibility of unauthorized use through sniffing or guessing. This risk is mitigated by support for challenge-response authentication \cite{just}. Specifically, the administrator first sends a short message to the device requesting access, and then the device sends back a challenge message containing a nonce string. The administrator then sends back a response message containing a hash value of the password and nonce string, in addition to a command, and the device executes the command only if the hash value is correct. If an attacker requests access to the device using a short message with a spoofed origin as the administrator's phone number, the device sends back a challenge message to the administrator; therefore, the attacker cannot generate a correct hash value. This allows the device to reject malicious commands received via short messages that do not contain the correct hash value.

OMA DM, a device management protocol that uses SMS, requires authentication that is not based on the origin of the SMS. The OMA DM security specification defines two authentication schemes: the basic and MD5 digest schemes \cite{omadm2}. The basic scheme authenticates the sender based on a Base64-encoded string that is a concatenation of the sender’s user ID and password. However, it must be noted that this scheme is vulnerable to mobile network sniffing \cite{solnik}. In contrast, the MD5 digest scheme authenticates the sender based on the MD5 hash of the nonce string issued by the authenticator in combination with the sender’s credentials. Additionally, four optional authentication schemes are allowed. Therefore, using OMA DM for the remote management of IoT devices is an effective mitigation measure against authentication bypass via SMS origin spoofing.

\subsection{Limitations}

This study demonstrated the threat against only one product; it did not demonstrate attacks that exploit various actions of different products. In this study, the threat was demonstrated by bypassing the originating number-based authentication for SMS-based remote management implemented in a product from Teltonika Networks, and a threat scenario was presented that exploited the action to disable IP blocking. Therefore, verification using other products and analysis of more serious threats using the SMS-based remote management actions they provide are beyond the scope of this study. Threats that exploit the various actions provided by IoT devices are limited to a discussion mentioned before, and further verification is a subject for future research.

Because this study focuses on cellular IoT gateways, the threats posed to other IoT devices and machines have not been demonstrated. This study was conducted using cellular IoT gateways, which are commonly used in environments that require SMS-based remote management. Therefore, the threat posed by SMS origin spoofing to other IoT devices and machines is a subject for future research. In the future, AI-powered IoT devices and machines may fall victim to phishing techniques that are effective against humans. If SMS were used in such cases, the threat of exploitation would be more serious and damaging to our lives than the extent discussed in this study. Therefore, research focusing on such threats is worthwhile.

The threats posed to IoT devices that use protocols other than SMS for remote management remain unclear. More recently, wireless communication with IoT devices has been developing communication standards for realizing low-power wide-area networks (LPWAN). Considering such developments, LTE-M and NB-IoT have been developed by 3GPP as wireless communication standards for IoT devices over mobile networks \cite{3gpp}. Given that these protocols have been implemented in 180 MNOs by 2023 \cite{gsa}, they are expected to replace SMS for the remote management of IoT devices in the future. Further research is required to clarify whether origin spoofing is feasible under such protocols and the threats it poses.

\section{Ethical Considerations}

All verifications in this study were performed under administrative control. The author owned all the devices and phone numbers used in the verifications; hence, no attempt was made to bypass the authentication of another person’s device. In addition, all verifications were performed using devices approved by the Japanese Radio Law. The results obtained from the verification have already been reported to all cellular IoT gateway vendors providing originating number-based authentication, and mitigation measures are being considered by each vendor. The specifics of the threat demonstration have been reported to Teltonika Networks; however, no mitigation measures have been implemented thus far because they enable password-based authentication by default.

\section{Conclusion}

This study revealed that SMS origin spoofing can be exploited to bypass the authentication of IoT devices. The focus of this study was on IoT devices that authenticate remote management commands based on the originating number of an SMS. Accordingly, the possibility of authentication bypass via SMS origin spoofing was investigated. Specifically, the specifications of major cellular IoT gateways were evaluated with a focus on SMS-based remote management. Moreover, the hypothesis of an authentication bypass was verified. The results showed that 25 of the 32 targeted products support SMS-based remote management and 20 of them implement originating number-based authentication. Furthermore, one product was demonstrated to be remotely exploitable through authentication bypassing. This proves that SMS origin spoofing not only threatens text communication between people but also puts machine communication at risk. Future research should focus on verifying this hypothesis using other IoT devices and machine communication protocols to completely mitigate the revealed threats.

\bibliographystyle{ACM-Reference-Format}
\bibliography{base}

\end{document}